\documentclass[twocolumn,nofootinbib,fleqn]{revtex4-1}
\usepackage{tikz}
\usepackage{amsmath, amsthm, amssymb}
\usepackage{slashed}
\usepackage{url}
\usepackage[pdftex]{hyperref}
\usepackage{xcolor}
\usepackage{bm,mathrsfs}
\usepackage{mathtools}
\usepackage{subfigure}
\usepackage{siunitx}
\usepackage{cleveref}

\newcommand{\diff}{\mathop{}\!\mathrm{d}} 
\newcommand{\GeV}{\giga\electronvolt}
\newcommand{\TeV}{\tera\electronvolt}
\newcommand{\mydash}{\raisebox{2pt}{\rule{4pt}{0.2pt}}}

\begin{document}
\tikz[overlay,remember picture]{
\node[anchor=north east] at 
([xshift=-2cm,yshift=-1cm]current page.north east){UCI\mydash HEP\mydash TR\mydash 2019\mydash 06};}
\preprint{UCI-HEP-TR-2019-06}

\title{Baryogenesis From Flavon Decays}
\author{Mu--Chun Chen}
\author{Seyda Ipek}
\author{Michael Ratz}
\affiliation{Department of Physics and Astronomy, University of
California, Irvine, CA 92697\mydash4575 USA}

\date{\today}
\begin{abstract}
Many popular attempts to explain the observed patterns of fermion masses involve
a flavon field. Such weakly coupled scalar fields tend to dominate
the energy density of the universe before they decay. If the flavon decay
happens close to the electroweak transition, the right\mydash handed electrons stay
out of equilibrium until the sphalerons shut off. We show that an asymmetry in
the right\mydash handed charged leptons produced in the decay of a flavon can explain
the baryon asymmetry of the universe.
\end{abstract}

\maketitle

\section{Introduction}
\label{sec:intro}
There is more matter than antimatter in the universe. This baryon asymmetry of
the universe (BAU) is measured to be \cite{Ade:2015xua, Riemer-Sorensen:2017vxj}
\begin{align}
\eta_\mathrm{obs}\equiv \frac{n_B-n_{\bar{B}}}{s}\simeq 8\times 10^{-11}\;,
\end{align}
where $n_{B(\bar{B})}$ is the (anti-)baryon number density and $s$ is the entropy density.

In order to explain the BAU, the three so\mydash called Sakharov conditions
\cite{Sakharov:1967dj} must be satisfied:
\textbf{(i)} Baryon (or lepton) number must be violated, \textbf{(ii)} C and CP
must be violated and \textbf{(iii)} There must be departure from thermal
equilibrium. 

Baryon number and CP are violated in the electroweak (EW) sector of the Standard Model (SM). 
However the CP violation in the SM is too small to account for the BAU. In addition, there
is no out\mydash of\mydash equilibrium process in the SM. In order to produce the BAU, beyond
the SM physics with additional degrees of freedom is needed. These degrees of freedom can also solve  
other deficiencies in the SM. Most notably example is in leptogenesis~\cite{Fukugita:1986hr}, 
where the right\mydash handed neutrinos are the only required additional degrees of freedom. 
These right\mydash neutrinos are part of the see\mydash saw mechanism~\cite{Minkowski:1977sc}, 
and thus providing an explanation for the smallness of neutrino masses. Through their  
out\mydash of\mydash equilibrium decays, the observed BAU can be obtained.

In this Letter, we show that another degree of freedom, the \emph{flavon}, that conceivably plays an
important role in understanding fermion masses, can also
generate the observed BAU. In general, a flavon is a scalar particle whose
vacuum expectation value (VEV) dictates the (Yukawa) couplings of the SM
fermions, with  the most popular example being the Froggatt--Nielsen
field~\cite{Froggatt:1978nt}. We show that the out\mydash of\mydash equilibrium
decays of such a flavon can be the necessary beyond\mydash the\mydash SM physics
that generates the observed BAU. The flavon decays may create a left\mydash
right charged lepton asymmetry. In a universe where the flavon dominates the
energy density, the right\mydash handed electrons may not come into chemical
equilibrium, and an asymmetry in these right\mydash handed electron asymmetry
can be converted into a baryon number asymmetry by sphalerons.

\section{Basic mechanism}
\label{sec:model}

Consider the couplings of the SM charged leptons to a SM singlet scalar, the flavon $S$, charged under the flavor symmetry 
$\mathrm{U(1)}_\mathrm{FN}$,
\begin{align}\label{eq:FlavonCoupling1}
 \mathscr{L}&\supset y_0^{fg}\,\left(\frac{v_S+S}{\Lambda}\right)^{n_{fg}}
 \overline{e}_\mathrm{R}^g \cdot\phi^*\cdot\ell_\mathrm{L}^f +\text{h.c.}
\end{align}
Here, $\phi$ denotes the electroweak Higgs doublet, $\ell_\mathrm{L}^f$ is the
$f^\mathrm{th}$\mydash generation left\mydash handed lepton doublet, 
and $e_\mathrm{R}^g$ represents the $g^\mathrm{th}$\mydash generation right\mydash handed charged lepton.
$v_{S}$ is the flavon VEV, with $\Lambda$ being the
cutoff scale of the flavor symmetry, $y_{0}$ an $\mathcal{O}(1)$ coupling constant 
and $n_{fg}$  integers related to the Froggatt--Nielsen charges under $\mathrm{U(1)}_\mathrm{FN}$.
In the following, we will restrict ourselves to the simplest version of the scenario, in which 
the flavon only couples to leptons. As a benchmark model, we
will take 
\begin{align}
y_0\simeq1\;,\quad \varepsilon=\frac{v_S}{\Lambda}=0.2\;,\quad
n_e=9\;,\quad n_\tau=3\;.
\end{align}

The key observation in our scenario is that the flavon decays 
\begin{align}\label{eq:flavondecays}
 S &\to\bar\ell_\mathrm{L}+\phi+ e_\mathrm{R}\;,&
 S^\ast &\to \ell_\mathrm{L}+\phi^\ast+\bar{e}_\mathrm{R}\;, 
\end{align}
are left\mydash right violating, e.g.\ an $S$ particle only decays into
$\ell_\mathrm{L}$ antiparticle and $e_\mathrm{R}$ particle.  Crucially, if there
is an initial flavon asymmetry, say an excess of $S$ over $S^{\ast}$, after the
flavons decay, there will be more left\mydash handed antileptons than
left\mydash handed leptons. Note that the total lepton number, $L$, is
conserved because the asymmetry in the left\mydash handed leptons will be
balanced by an equal and opposite asymmetry in the right\mydash handed leptons.
However, sphalerons will only act on left\mydash handed antileptons and
partially convert this asymmetry into a baryon ($B$) asymmetry. This is
similar to Dirac leptogenesis~\cite{Dick:1999je}, in which nonzero baryon number
asymmetry is possible  with $B\!-\!L$ conservation, as it utilizes the
features that sphalerons only couple to  left\mydash handed fields.  In the
scenario discussed here, the left\mydash right asymmetry originates from a
primordial flavon\mydash antiflavon asymmetry, which we will discuss below in
more detail.  

\section{Flavon Asymmetry}
\label{sec:Sasym}

Our scenario requires the flavon to have a large asymmetry when it decays. This
means that an asymmetry must be created and that flavon number must be conserved during the flavon oscillations. It has been pointed out in the context of
the Affleck--Dine scenario~\cite{Affleck:1984fy} that scalar field dynamics in
the early universe can exhibit these features. As explained in
\cite[Section~III.C]{Dine:2003ax}, in supersymmetric theories the mass terms of
scalar fields $S$ preserve $S$\mydash number while interaction terms do not. These
interactions may provide an ``initial kick'' (in the terminology of
\cite{Kitano:2008tk}) which leads to an angular momentum of the $S$ field in the
complex plane, i.e.\ an $S$\mydash number asymmetry. These terms become unimportant
later when the $S$ field performs coherent oscillations, such that the asymmetry persists until the flavon decays. 

A reason for concern might be the origin of the $\mathrm{U}(1)$ symmetry that ensures flavon number conservation. At
first sight, one may conclude that if $\mathrm{U}(1)_\mathrm{FN}$ gets broken
spontaneously, one is left with two options. \textbf{(i)}
$\mathrm{U}(1)_\mathrm{FN}$ is local, in which case the would\mydash be
Goldstone mode gets eaten, and one is left with a real scalar and no
$\mathrm{U}(1)_S$ that allows us to define the $S$\mydash number. \textbf{(ii)}
$\mathrm{U}(1)_\mathrm{FN}$ is global, in which case there is a Goldstone mode,
and the situation is even worse. However, in explicit models the situation is often
richer than that. For instance, in supersymmetric scenarios, the scalar fields
are complex. In such models, one starts with more degrees of freedom and the
flavon is a complex linear combination of these fields, whose mass term preserves
a $\mathrm{U}(1)$ symmetry. There are also non\mydash supersymmetric examples that
have additional custodial symmetries (see
e.g.~\cite{Hambye:2008bq,Arcadi:2016kmk} for recent applications of these
symmetries in dark matter model building). We will assume that for energy scales
far below $\Lambda$ an approximate $\mathrm{U}(1)_S$ is preserved by the flavon
potential, i.e.\
\begin{equation}
 \mathscr{V}_S=m^2|S|^2+\left(\begin{array}{@{}l@{}}
 \text{$S$\mydash number violating terms}\\ 
\text{suppressed by powers of $\Lambda$}\end{array} \right)\;.
\end{equation}

\section{Flavon Cosmoloy}
\label{sec:flavondecay}

By assumption, the flavon we consider is a weakly coupled scalar field. Hence it can perform
coherent oscillations around the $T=0$ minimum of its potential. In fact,
thermal corrections to the potential will push the flavon away from its
expectation value at $T=0$~\cite{Lillard:2018zts}. The energy density stored in
these oscillations, $\rho_S$, only drops as $a^{-3}$ whereas the energy density of
radiation, $\rho_\mathrm{rad}$, drops as $a^{-4}$, where $a$ is the scale
factor.  It is expected that at a time $t_{\ast}$, corresponding to a
temperature $T_{\ast}$, the energy density stored in flavon oscillations starts
dominating over the radiation contribution. We define this point as 
\begin{equation}
 \rho_S|_{_{T=T_\ast}}=\rho_\mathrm{rad}|_{_{T=T_\ast}}\;.
\end{equation}

The flavon decays to SM leptons with a decay rate 
\begin{align}
 \Gamma_S\sim\frac{1}{\varepsilon}\,
 \frac{|n_\tau\,y_\tau|^2}{64\pi^3}\,\frac{m_S^3}{\Lambda^2}\;.  
 \label{eq:Gflavon}
\end{align}
Most of the decay products thermalize with the radiation and contribute to the
radiation density. However, as we will show below, the right\mydash handed
electrons might not come into chemical equilibrium before the sphalerons have
switched off. 

The evolution of the relevant energy densities is given by 
\begin{subequations}\label{eq:Edensity}
\begin{align}
\frac{\diff\rho_S}{\diff t}+3 H\,\rho_S&= -\Gamma_S\,\rho_S\;,  \\
\frac{\diff\rho_\mathrm{rad}}{\diff t}+4 H\,\rho_\mathrm{rad} &= 
\Gamma_S\,\rho_S\;,
\end{align}
\end{subequations}
where the Hubble rate is determined by the Friedmann equation
\begin{align}
 H^2 &= \frac{8\pi}{3M_\mathrm{Pl}^2}(\rho_S+\rho_\mathrm{rad})\;, \label{eq:Hubble}
\end{align}
with $M_\mathrm{Pl}\simeq \SI{1.2e19}{\GeV}$ being the Planck mass. 

\begin{figure}[h]
\includegraphics[scale=.4]{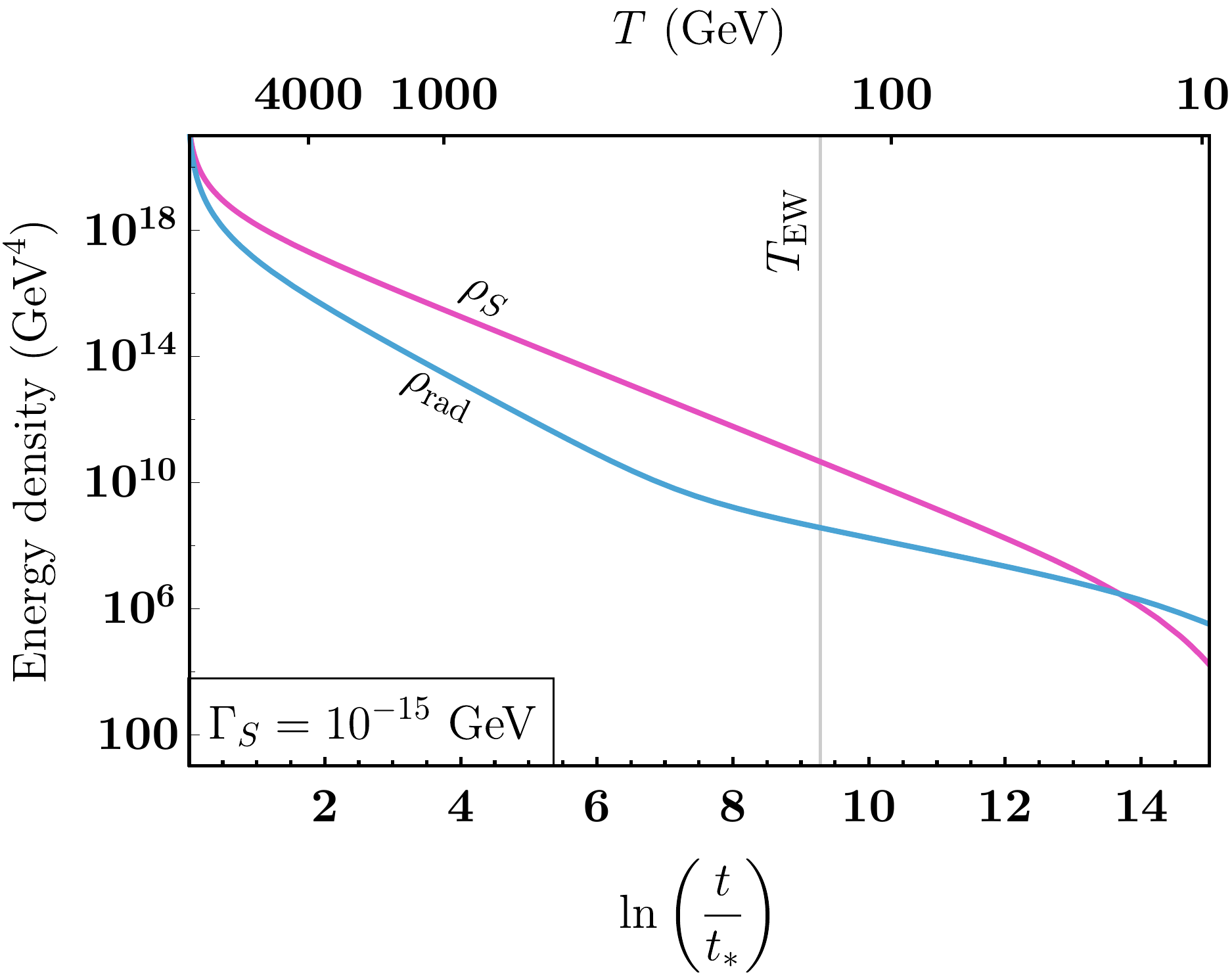}
\caption{Energy densities stored in flavon oscillations, $\rho_S$, and in
radiation $\rho_{\rm rad}$, with respect to temperature. As a benchmark scenario,
it is assumed that $\rho_S$ starts dominating at $T_\ast=\SI{100}{\TeV}$ and the
flavon decay rate is $\Gamma_S= \SI{e-15}{\GeV}$, which corresponds to
a decay temperature $T_\mathrm{d}\simeq \SI{10}{\GeV}$. We also show the temperature when the EW sphalerons shut off, $T_{\rm EW}\sim160~$GeV.}\label{fig:Edensity}
\end{figure}

 We solve this set of equations numerically and show a benchmark scenario in
\Cref{fig:Edensity}. Analytical approximations for the two energy densities can
be given at a time $t$ before the flavon decays, $t_\ast<t<\tau\sim \Gamma_S^{-1}$, as
\begin{subequations}
\begin{align}
 \rho_S (t)
 &\simeq\frac{M_{\rm Pl}^2}{6\pi t^2}\,\mathrm{e}^{-\Gamma_S\,t}\;, \\
 \rho_\mathrm{rad}(t)&\simeq\frac{M_{\rm Pl}^2 t_\ast^{2/3}}{6\pi\,t^{8/3}}+ \frac{\Gamma_S\,M_{\rm Pl}^2}{10\pi\,t}\;.
\end{align}
\end{subequations}
The first term in $\rho_\mathrm{rad}$ is the initial radiation energy
density and the second term is generated from flavon decays. This second
term starts taking over $\rho_\mathrm{rad}$ long before $\Gamma_S^{-1}$. 

We will show in the next section that BAU can be produced via flavon decays for
$\Gamma_S\sim O(10^{-13}-10^{-17}~{\rm GeV})$. Thus, using \Cref{eq:Gflavon} we will require the flavon mass to be
\begin{align}
\frac{m_S}{\si{\TeV}} &\sim
\left(\frac{\Lambda}{\SI{e9}{\GeV}}\right)^{\frac{2}{3}}
\left(\frac{\Gamma_S}{\SI{e-15}{\GeV}}\right)^{\frac{1}{3}}\;, \label{eq:mS}
\end{align}
such that $\Lambda$ is below the Planck scale.

\section{Generation and non--equilibration of right--handed electrons}
\label{sec:RHelectrons}

Among other leptons, flavon decays into right\mydash handed electrons with a branching
fraction of $B_e\sim \left(\frac{n_e\,y_e}{n_\tau\,y_\tau}\right)^2\sim 7.5\times
10^{-7}$. Through its decays, the flavon asymmetry will get partially converted
into an asymmetry in right\mydash handed electrons. Like in
leptogenesis~\cite{Fukugita:1986hr}, this asymmetry is turned into a baryon
asymmetry by sphalerons. Similarly to Dirac leptogenesis~\cite{Dick:1999je}, our
scenario does not require $B\!-\!L$ violation. However, our scenario works both
for Dirac and Majorana neutrinos.

\begin{figure}[h]
\includegraphics[scale=.4]{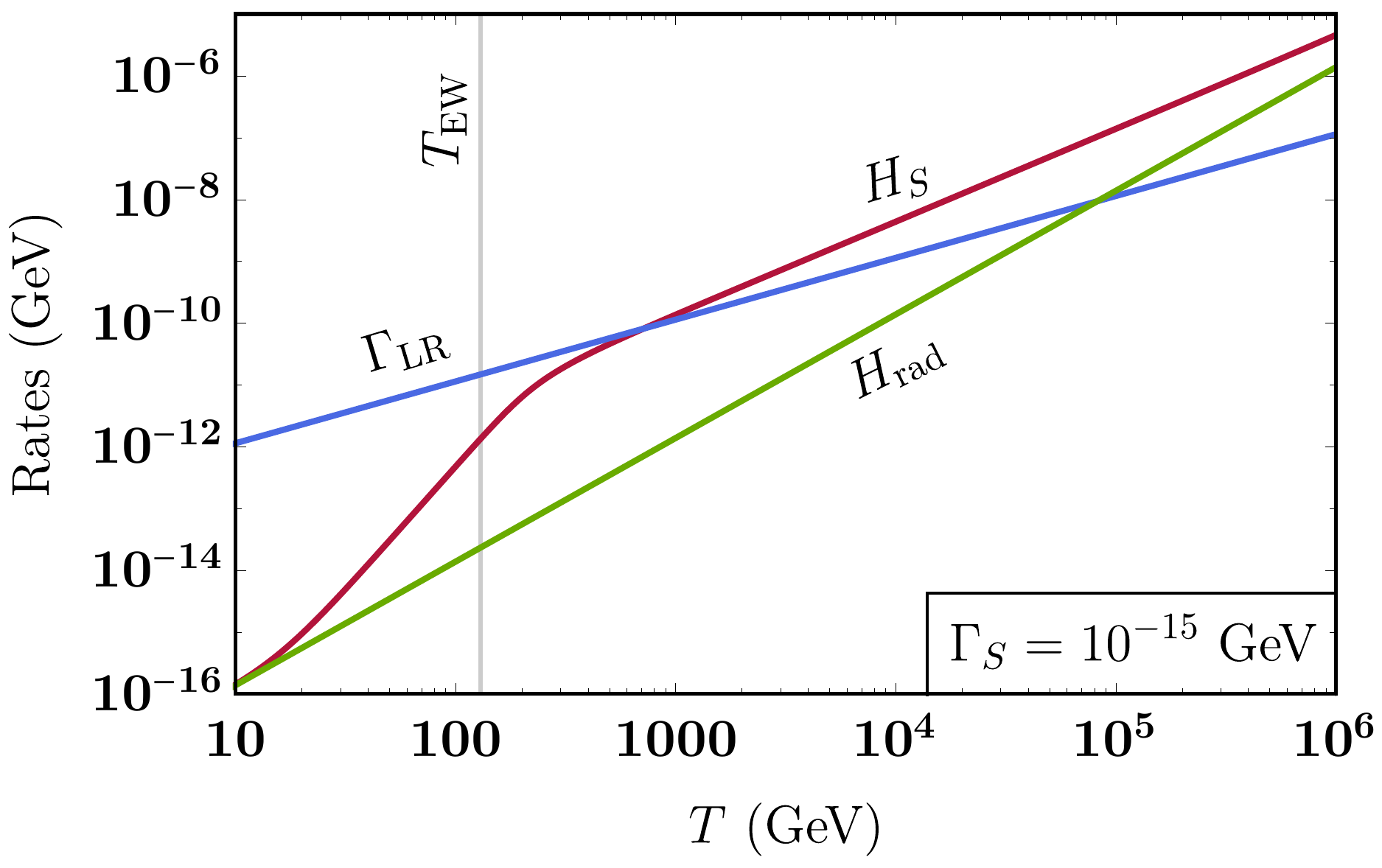}
\caption{Comparison of the equilibration rate of right\mydash handed electrons,
$\Gamma_\mathrm{LR}$, to the Hubble rates in a radiation dominated universe,
$H_\mathrm{rad}$ and in a universe with intermediate flavon domination, $H_S$.
For this benchmark case the flavon energy density starts dominating at $T_\ast
=\SI{e7}{\TeV}$ and the flavon decays at
$T_\mathrm{d}\simeq\SI{10}{\GeV}$.}\label{fig:Rates}
\end{figure}

Right\mydash handed electrons equilibrate with the SM plasma mainly through their
interactions with the Higgs boson and $2\to2$ scatterings. This equilibration
rate has been recently calculated to be~\cite{Bodeker:2019ajh}
\begin{align}
\Gamma_\mathrm{LR}&\simeq10^{-2}\,y_e^2\,T\;,\label{eq:GammaLR}
\end{align}
which is larger by almost an order of magnitude  than the initial
estimate~\cite{Campbell:1992jd} and by a factor of a few than a
refined estimate~\cite{Cline:1993bd}. Comparing this rate to the Hubble rate for
a radiation\mydash dominated universe, one finds that the right\mydash handed electrons come
into equilibrium at $T\sim \SI{e5}{\GeV}$. Hence, in standard cosmology, any
asymmetry in the right\mydash handed electrons would be washed out long before the EW transition at $T\sim160~$GeV. However, in a
universe that is dominated by a flavon until temperatures around the electroweak 
scale, right\mydash handed electrons may \emph{not} equilibrate. (See \Cref{fig:Rates}.)

We find the asymmetry in the number density of right\mydash handed electrons after the
flavon decays by solving the Boltzman equation for $n_\mathrm{R} =
n_{e_\mathrm{R}}-n_{\bar{e}_\mathrm{R}}$,
\begin{align}
\frac{\diff n_\mathrm{R}}{\diff t}=-&3 H\,n_\mathrm{R} - 
\Gamma_{\rm LR}\,n_\mathrm{R}+B_e\,\Gamma_S\, n_S \;, \label{eq:dnRdt}
\end{align}
where $B_e$ is the flavon branching ratio to electrons The asymmetry in the flavon number density, $n_S$, is parametrized through an initial flavon asymmetry, $\eta_S$, as
\begin{align}
 n_S&\equiv\eta_S\,\frac{\rho_S}{m_S}\;. \label{eq:nS}
\end{align}

\begin{figure}[h]
\includegraphics[scale=.4]{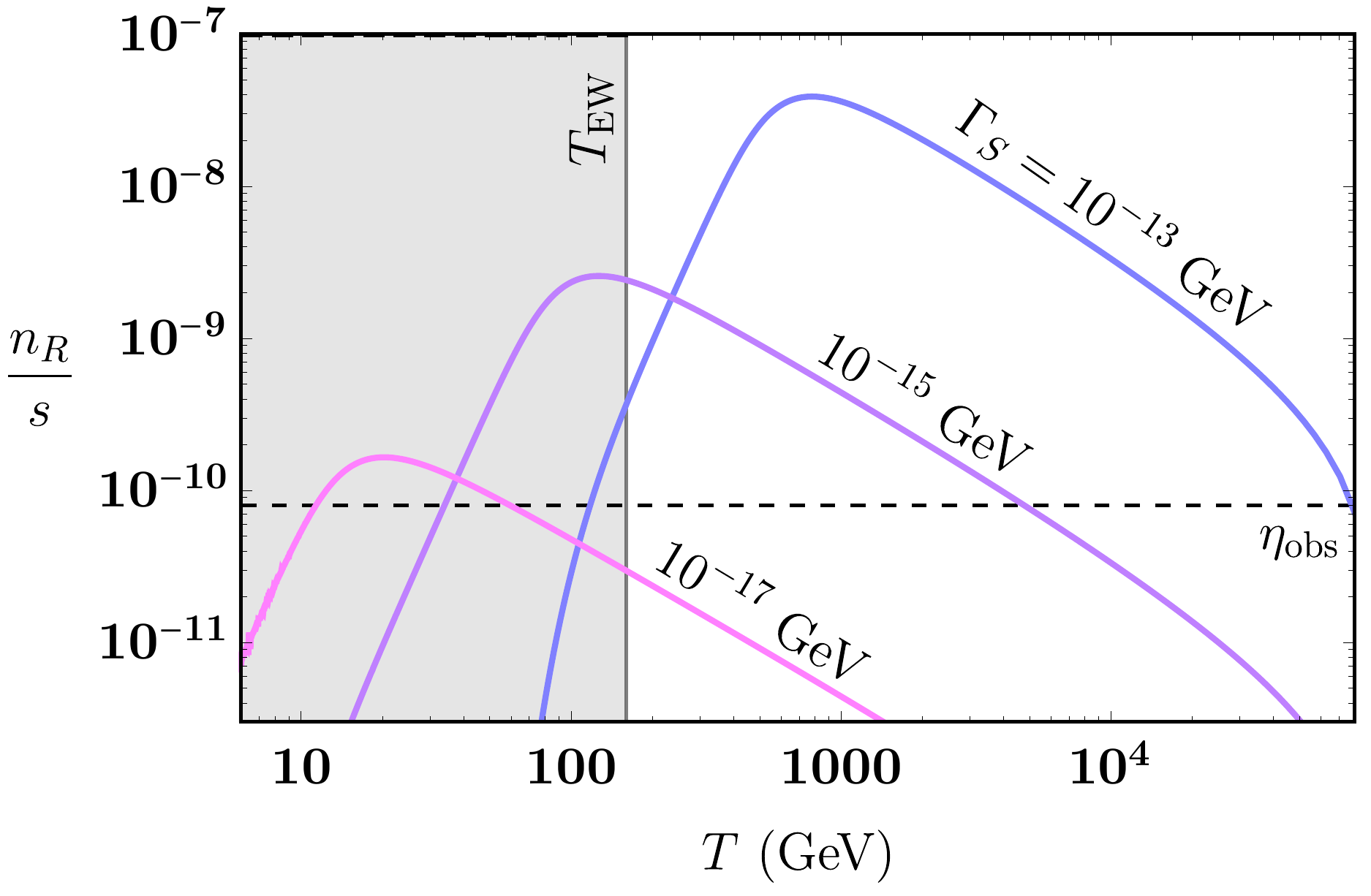}
\caption{Right\mydash handed electron asymmetry generated through asymmetric flavon
decays for different values of the flavon decay rate $\Gamma_S$. Observed baryon asymmetry $\eta_{\rm obs}\simeq 8\times 10^{-11}$ is shown for reference.}\label{fig:eRdensity}
\end{figure}

We solve for $n_\mathrm{R}$ together with \Cref{eq:Edensity}. In
\Cref{fig:eRdensity} we show the dimensionless quantity
$\eta_\mathrm{R}=\frac{n_\mathrm{R}}{s}$, where $s$ is the entropy density, for
a benchmark scenario with 
\begin{align}
 m_S=\SI{1}{\TeV}\;,~\eta_S=1\;,~T_\ast=\SI{100}{\TeV}\;.
\end{align}
As usual, the temperature is defined as the temperature of radiation, which may
or may not dominate the energy density of the universe.

The behavior of $\eta_R$ can be understood qualitatively as follows. A
right\mydash handed electron asymmetry is produced through flavon decays and is
proportional to the flavon number density. When the Hubble rate drops below the
left\mydash right equilibration rate, the asymmetry is washed out by SM
interactions. The time when this washout starts depends on the flavon lifetime,
$\tau=\Gamma_S^{-1}$, and is generally before $\tau$. Smaller $\Gamma_S$ can
delay the equilibration until after sphalerons shut off at $T_{\rm EW}\sim
160$~GeV. However, this also means that a smaller number of flavons decay before
$T_{\rm EW}$. Hence, in general, there is balancing between the flavon lifetime and
$\Gamma_{\rm LR}$ that describes the region where the right amount of
right\mydash handed electron asymmetry is produced before the sphalerons shut
off.

\section{Baryon Asymmetry}
\label{sec:bau}

\begin{figure}
\includegraphics[scale=.4]{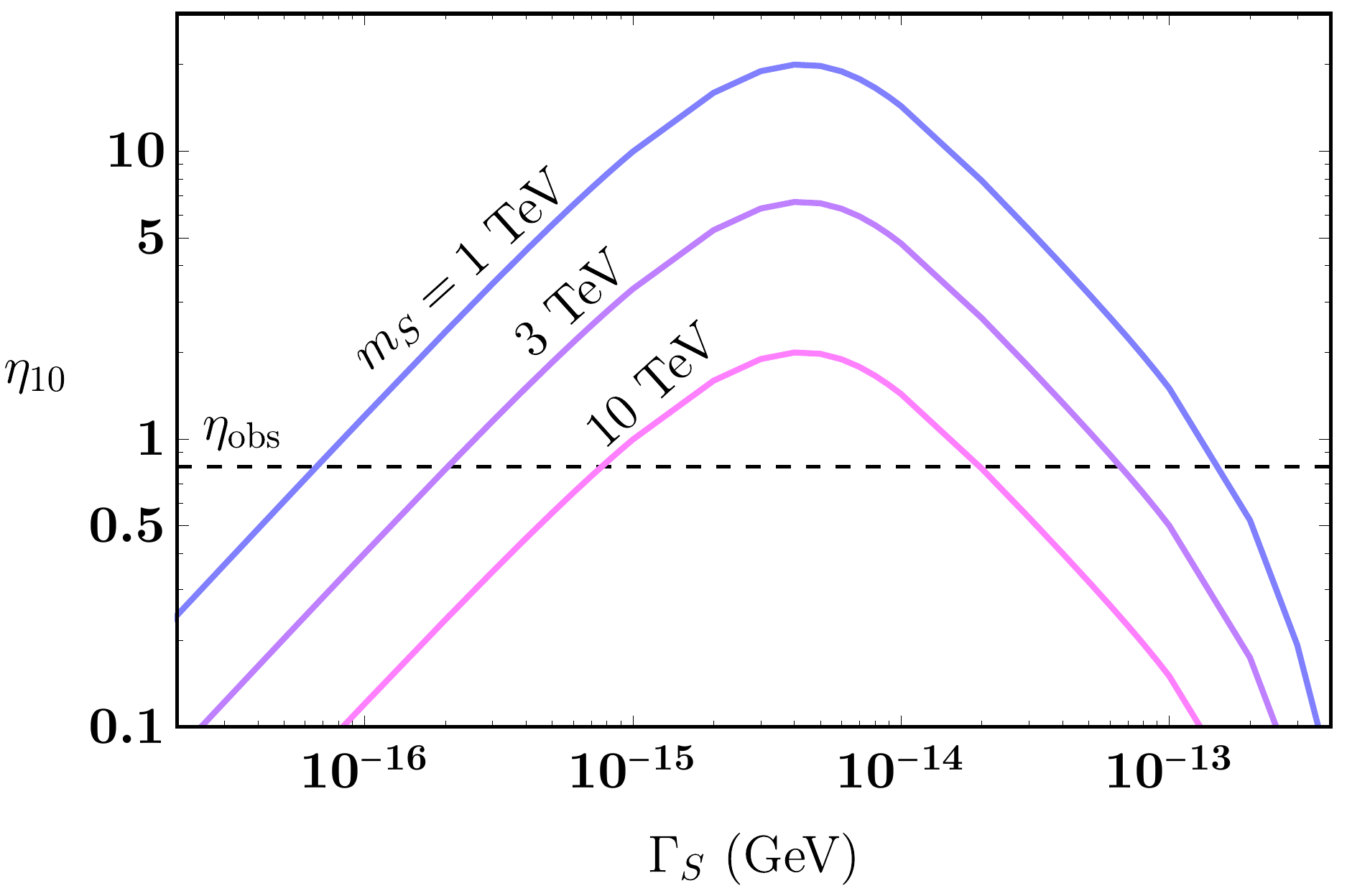}
\caption{Baryon asymmetry ($\eta_{10}\equiv \eta_B\times 10^{10}$) of the universe as a function of the flavon decay
rate $\Gamma_S$ for different values of the flavon mass $m_S$. As a benchmark scenario we take  the
initial flavon asymmetry $\eta_S=1$. The observed baryon asymmetry 
$\eta_\mathrm{obs}\simeq 8\times10^{-11}$ is shown for reference.}\label{fig:etaBvGS}
\end{figure}

In the SM $B\!+\!L$ number is violated by EW processes called
sphalerons~\cite{Klinkhamer:1984di,Kuzmin:1985mm}. Sphalerons are unsuppressed
above the EW transition temperature $T_\mathrm{EW}$, with a thermal rate
$\Gamma_\mathrm{sph}\simeq (\alpha_2 T)^4$, while they are exponentially
suppressed due to finite gauge boson masses after EW symmetry breaking. While
sphalerons violate $B\!+\!L$, they conserve $B\!-\!L$. Hence a lepton asymmetry
will be turned into a baryon asymmetry \cite{Fukugita:1986hr}. In our scenario this baryon
asymmetry is given by~\cite{Cline:1993bd} 
\begin{align}
 \eta_B &\equiv
\frac{n_B}{s}\simeq\frac{198}{481}\frac{n_\mathrm{R}}{s}\left|_{_{T_\mathrm{EW}}}\right.~.
\end{align}

We numerically solve \Cref{eq:Edensity,eq:dnRdt} to find the electron asymmetry
and the entropy of the universe at $T_\mathrm{EW}\simeq\SI{160}{GeV}$. Our
results are shown in \Cref{fig:eRdensity,fig:etaBvGS}. As can be seen in \Cref{fig:etaBvGS}, there
is a large parameter space where asymmetric flavon decays produce the observed
BAU. We comment on the dependence of $\eta_B$ on the parameters of the model. 
\begin{itemize}
\item Our scenario is rather sensititve to the flavon decay
 rate. If the flavon decays too early, $T_\mathrm{d}\gg T_\mathrm{EW}$, 
 right\mydash handed electrons equilibrate and (essentially) no baryon asymmetry
 is generated. If the flavon decays too late, $T_\mathrm{d}\ll T_\mathrm{EW}$,
 the right\mydash handed electron asymmetry is produced when sphalerons are
 inoperative, and again no baryon asymmetry emerges. We find that the observed
 baryon asymmetry is produced for $10^{-16}\,\mathrm{GeV}\lesssim \Gamma_S
 \lesssim 10^{-13}\,\mathrm{GeV}$.

\item The baryon asymmetry is virtually independent of the time $t_\ast$ when
 the flavon domination starts as long as $T_\ast \gg
 T_\mathrm{d},T_\mathrm{EW}$.

\item As the flavon mass increases, its number density drops. This is because
 each flavon decay produces entropy, the dilution factor being
 $m_S/T_\mathrm{d}$. Hence the baryon asymmetry is inversely proportional to
 $m_S$. The flavon decay rate also depends on its mass. For given values of
 $m_S$ and $\Gamma_S$, the appropriate flavon scale  $\Lambda$ can be found from
 \Cref{eq:mS}. We find that a flavon with $\mathcal{O}(\mathrm{TeV})$ mass can produce
 the observed BAU. This requires $\Lambda\sim \mathcal{O}(10^{9}\,\mathrm{GeV})$. We
 note that for such a large flavon scale, constraints on the flavon mass is
 $\mathcal{O}(100\,\mathrm{GeV})$~\cite{Bauer:2016rxs}.
 
\item The baryon asymmetry is directly proportional to the primordial flavon
 asymmetry $\eta_S$. We show our results in \Cref{fig:etaBvGS} for the maximal
 value of $\eta_S=1$. If we allow for a flavon as light as 200~GeV, we only
 require $\eta_S\gtrsim 10^{-2}$.

\end{itemize}

\section{Conclusions}

In this Letter, we have shown that the observed baryon asymmetry can be produced
in flavon decays. This scenario requires the flavon  to decay around the time of
the electroweak transition and that the flavon carries an 
$\mathcal{O}(10^{-2}-1)$ primordial asymmetry. The flavon mass is
$\mathcal{O}(1-10\,\mathrm{TeV})$. The viable parameter ranges are illustrated
in \Cref{fig:etaBvGS}. The role of the flavon is twofold: 
\begin{enumerate}
 \item Its decays produce a left\mydash right asymmetry in the lepton sector, the
  left\mydash handed part of which is converted to a baryon asymmetry by sphalerons.
 \item It dominates the universe before the EW scale, thus increasing the
  Hubble rate and preventing the right\mydash handed electrons from equilibrating.
\end{enumerate}

Our scenario is rather constrained and thus predictive. We ``lose'' a
significant amount of efficiency because our scalar is a flavon, i.e.\ it
predominantly decays into the heavier generations. A more ``efficient'' version
of this scenario would be to couple a scalar to light fermions only, in which
case one may also consider the decay into up or even down quarks. However, it is
arguably an appealing aspect of our scenario that we do not have to introduce a
new degree of freedom in order to explain the observed baryon asymmetry. Rather,
a traditional flavon can do that without being tweaked. 

\section*{Acknowledgements}
This work is supported in part by NSF Grant No.~PHY\mydash1620638 and by NSF
Grant No.~PHY\mydash1719438. SI acknowledges support from the UC Office of the
President via a  UC Presidential Postdoctoral fellowship. This work was
performed in part at Aspen Center for Physics,  which is supported by National
Science Foundation grant PHY\mydash1607611. 

\bibliography{ref}

\begin{thebibliography}{19}%
\makeatletter
\providecommand \@ifxundefined [1]{%
 \@ifx{#1\undefined}
}%
\providecommand \@ifnum [1]{%
 \ifnum #1\expandafter \@firstoftwo
 \else \expandafter \@secondoftwo
 \fi
}%
\providecommand \@ifx [1]{%
 \ifx #1\expandafter \@firstoftwo
 \else \expandafter \@secondoftwo
 \fi
}%
\providecommand \natexlab [1]{#1}%
\providecommand \enquote  [1]{``#1''}%
\providecommand \bibnamefont  [1]{#1}%
\providecommand \bibfnamefont [1]{#1}%
\providecommand \citenamefont [1]{#1}%
\providecommand \href@noop [0]{\@secondoftwo}%
\providecommand \href [0]{\begingroup \@sanitize@url \@href}%
\providecommand \@href[1]{\@@startlink{#1}\@@href}%
\providecommand \@@href[1]{\endgroup#1\@@endlink}%
\providecommand \@sanitize@url [0]{\catcode `\\12\catcode `\$12\catcode
  `\&12\catcode `\#12\catcode `\^12\catcode `\_12\catcode `\%12\relax}%
\providecommand \@@startlink[1]{}%
\providecommand \@@endlink[0]{}%
\providecommand \url  [0]{\begingroup\@sanitize@url \@url }%
\providecommand \@url [1]{\endgroup\@href {#1}{\urlprefix }}%
\providecommand \urlprefix  [0]{URL }%
\providecommand \Eprint [0]{\href }%
\providecommand \doibase [0]{http://dx.doi.org/}%
\providecommand \selectlanguage [0]{\@gobble}%
\providecommand \bibinfo  [0]{\@secondoftwo}%
\providecommand \bibfield  [0]{\@secondoftwo}%
\providecommand \translation [1]{[#1]}%
\providecommand \BibitemOpen [0]{}%
\providecommand \bibitemStop [0]{}%
\providecommand \bibitemNoStop [0]{.\EOS\space}%
\providecommand \EOS [0]{\spacefactor3000\relax}%
\providecommand \BibitemShut  [1]{\csname bibitem#1\endcsname}%
\let\auto@bib@innerbib\@empty
\bibitem [{\citenamefont {Ade}\ \emph {et~al.}(2016)\citenamefont {Ade} \emph
  {et~al.}}]{Ade:2015xua}%
  \BibitemOpen
  \bibfield  {author} {\bibinfo {author} {\bibfnamefont {P.~A.~R.}\
  \bibnamefont {Ade}} \emph {et~al.} (\bibinfo {collaboration} {Planck}),\
  }\href {\doibase 10.1051/0004-6361/201525830} {\bibfield  {journal} {\bibinfo
   {journal} {Astron. Astrophys.}\ }\textbf {\bibinfo {volume} {594}},\
  \bibinfo {pages} {A13} (\bibinfo {year} {2016})},\ \Eprint
  {http://arxiv.org/abs/1502.01589} {arXiv:1502.01589 [astro-ph.CO]}
  \BibitemShut {NoStop}%
\bibitem [{\citenamefont {Riemer-S{\o}rensen}\ and\ \citenamefont
  {Jenssen}(2017)}]{Riemer-Sorensen:2017vxj}%
  \BibitemOpen
  \bibfield  {author} {\bibinfo {author} {\bibfnamefont {S.}~\bibnamefont
  {Riemer-S{\o}rensen}}\ and\ \bibinfo {author} {\bibfnamefont {E.~S.}\
  \bibnamefont {Jenssen}},\ }\bibfield  {booktitle} {\emph {\bibinfo
  {booktitle} {{Universe 2017, 3(2), 44}}},\ }\href {\doibase
  10.3390/universe3020044} {\bibfield  {journal} {\bibinfo  {journal}
  {Universe}\ }\textbf {\bibinfo {volume} {3}},\ \bibinfo {pages} {44}
  (\bibinfo {year} {2017})},\ \Eprint {http://arxiv.org/abs/1705.03653}
  {arXiv:1705.03653 [astro-ph.CO]} \BibitemShut {NoStop}%
\bibitem [{\citenamefont {Sakharov}(1967)}]{Sakharov:1967dj}%
  \BibitemOpen
  \bibfield  {author} {\bibinfo {author} {\bibfnamefont {A.~D.}\ \bibnamefont
  {Sakharov}},\ }\href {\doibase 10.1070/PU1991v034n05ABEH002497} {\bibfield
  {journal} {\bibinfo  {journal} {Pisma Zh. Eksp. Teor. Fiz.}\ }\textbf
  {\bibinfo {volume} {5}},\ \bibinfo {pages} {32} (\bibinfo {year} {1967})},\
  \bibinfo {note} {[Usp. Fiz. Nauk161,no.5,61(1991)]}\BibitemShut {NoStop}%
\bibitem [{\citenamefont {Fukugita}\ and\ \citenamefont
  {Yanagida}(1986)}]{Fukugita:1986hr}%
  \BibitemOpen
  \bibfield  {author} {\bibinfo {author} {\bibfnamefont {M.}~\bibnamefont
  {Fukugita}}\ and\ \bibinfo {author} {\bibfnamefont {T.}~\bibnamefont
  {Yanagida}},\ }\href {\doibase 10.1016/0370-2693(86)91126-3} {\bibfield
  {journal} {\bibinfo  {journal} {Phys. Lett.}\ }\textbf {\bibinfo {volume}
  {B174}},\ \bibinfo {pages} {45} (\bibinfo {year} {1986})}\BibitemShut
  {NoStop}%
\bibitem [{\citenamefont {Minkowski}(1977)}]{Minkowski:1977sc}%
  \BibitemOpen
  \bibfield  {author} {\bibinfo {author} {\bibfnamefont {P.}~\bibnamefont
  {Minkowski}},\ }\href {\doibase 10.1016/0370-2693(77)90435-X} {\bibfield
  {journal} {\bibinfo  {journal} {Phys. Lett.}\ }\textbf {\bibinfo {volume}
  {67B}},\ \bibinfo {pages} {421} (\bibinfo {year} {1977})}\BibitemShut
  {NoStop}%
\bibitem [{\citenamefont {Froggatt}\ and\ \citenamefont
  {Nielsen}(1979)}]{Froggatt:1978nt}%
  \BibitemOpen
  \bibfield  {author} {\bibinfo {author} {\bibfnamefont {C.~D.}\ \bibnamefont
  {Froggatt}}\ and\ \bibinfo {author} {\bibfnamefont {H.~B.}\ \bibnamefont
  {Nielsen}},\ }\href {\doibase 10.1016/0550-3213(79)90316-X} {\bibfield
  {journal} {\bibinfo  {journal} {Nucl. Phys.}\ }\textbf {\bibinfo {volume}
  {B147}},\ \bibinfo {pages} {277} (\bibinfo {year} {1979})}\BibitemShut
  {NoStop}%
\bibitem [{\citenamefont {Dick}\ \emph {et~al.}(2000)\citenamefont {Dick},
  \citenamefont {Lindner}, \citenamefont {Ratz},\ and\ \citenamefont
  {Wright}}]{Dick:1999je}%
  \BibitemOpen
  \bibfield  {author} {\bibinfo {author} {\bibfnamefont {K.}~\bibnamefont
  {Dick}}, \bibinfo {author} {\bibfnamefont {M.}~\bibnamefont {Lindner}},
  \bibinfo {author} {\bibfnamefont {M.}~\bibnamefont {Ratz}}, \ and\ \bibinfo
  {author} {\bibfnamefont {D.}~\bibnamefont {Wright}},\ }\href {\doibase
  10.1103/PhysRevLett.84.4039} {\bibfield  {journal} {\bibinfo  {journal}
  {Phys. Rev. Lett.}\ }\textbf {\bibinfo {volume} {84}},\ \bibinfo {pages}
  {4039} (\bibinfo {year} {2000})},\ \Eprint
  {http://arxiv.org/abs/hep-ph/9907562} {arXiv:hep-ph/9907562 [hep-ph]}
  \BibitemShut {NoStop}%
\bibitem [{\citenamefont {Affleck}\ and\ \citenamefont
  {Dine}(1985)}]{Affleck:1984fy}%
  \BibitemOpen
  \bibfield  {author} {\bibinfo {author} {\bibfnamefont {I.}~\bibnamefont
  {Affleck}}\ and\ \bibinfo {author} {\bibfnamefont {M.}~\bibnamefont {Dine}},\
  }\href {\doibase 10.1016/0550-3213(85)90021-5} {\bibfield  {journal}
  {\bibinfo  {journal} {Nucl. Phys.}\ }\textbf {\bibinfo {volume} {B249}},\
  \bibinfo {pages} {361} (\bibinfo {year} {1985})}\BibitemShut {NoStop}%
\bibitem [{\citenamefont {Dine}\ and\ \citenamefont
  {Kusenko}(2003)}]{Dine:2003ax}%
  \BibitemOpen
  \bibfield  {author} {\bibinfo {author} {\bibfnamefont {M.}~\bibnamefont
  {Dine}}\ and\ \bibinfo {author} {\bibfnamefont {A.}~\bibnamefont {Kusenko}},\
  }\href {\doibase 10.1103/RevModPhys.76.1} {\bibfield  {journal} {\bibinfo
  {journal} {Rev. Mod. Phys.}\ }\textbf {\bibinfo {volume} {76}},\ \bibinfo
  {pages} {1} (\bibinfo {year} {2003})},\ \Eprint
  {http://arxiv.org/abs/hep-ph/0303065} {arXiv:hep-ph/0303065 [hep-ph]}
  \BibitemShut {NoStop}%
\bibitem [{\citenamefont {Kitano}\ \emph {et~al.}(2008)\citenamefont {Kitano},
  \citenamefont {Murayama},\ and\ \citenamefont {Ratz}}]{Kitano:2008tk}%
  \BibitemOpen
  \bibfield  {author} {\bibinfo {author} {\bibfnamefont {R.}~\bibnamefont
  {Kitano}}, \bibinfo {author} {\bibfnamefont {H.}~\bibnamefont {Murayama}}, \
  and\ \bibinfo {author} {\bibfnamefont {M.}~\bibnamefont {Ratz}},\ }\href
  {\doibase 10.1016/j.physletb.2008.09.049} {\bibfield  {journal} {\bibinfo
  {journal} {Phys. Lett.}\ }\textbf {\bibinfo {volume} {B669}},\ \bibinfo
  {pages} {145} (\bibinfo {year} {2008})},\ \Eprint
  {http://arxiv.org/abs/0807.4313} {arXiv:0807.4313 [hep-ph]} \BibitemShut
  {NoStop}%
\bibitem [{\citenamefont {Hambye}(2009)}]{Hambye:2008bq}%
  \BibitemOpen
  \bibfield  {author} {\bibinfo {author} {\bibfnamefont {T.}~\bibnamefont
  {Hambye}},\ }\href {\doibase 10.1088/1126-6708/2009/01/028} {\bibfield
  {journal} {\bibinfo  {journal} {JHEP}\ }\textbf {\bibinfo {volume} {01}},\
  \bibinfo {pages} {028} (\bibinfo {year} {2009})},\ \Eprint
  {http://arxiv.org/abs/0811.0172} {arXiv:0811.0172 [hep-ph]} \BibitemShut
  {NoStop}%
\bibitem [{\citenamefont {Arcadi}\ \emph {et~al.}(2016)\citenamefont {Arcadi},
  \citenamefont {Gross}, \citenamefont {Lebedev}, \citenamefont {Mambrini},
  \citenamefont {Pokorski},\ and\ \citenamefont {Toma}}]{Arcadi:2016kmk}%
  \BibitemOpen
  \bibfield  {author} {\bibinfo {author} {\bibfnamefont {G.}~\bibnamefont
  {Arcadi}}, \bibinfo {author} {\bibfnamefont {C.}~\bibnamefont {Gross}},
  \bibinfo {author} {\bibfnamefont {O.}~\bibnamefont {Lebedev}}, \bibinfo
  {author} {\bibfnamefont {Y.}~\bibnamefont {Mambrini}}, \bibinfo {author}
  {\bibfnamefont {S.}~\bibnamefont {Pokorski}}, \ and\ \bibinfo {author}
  {\bibfnamefont {T.}~\bibnamefont {Toma}},\ }\href {\doibase
  10.1007/JHEP12(2016)081} {\bibfield  {journal} {\bibinfo  {journal} {JHEP}\
  }\textbf {\bibinfo {volume} {12}},\ \bibinfo {pages} {081} (\bibinfo {year}
  {2016})},\ \Eprint {http://arxiv.org/abs/1611.00365} {arXiv:1611.00365
  [hep-ph]} \BibitemShut {NoStop}%
\bibitem [{\citenamefont {Lillard}\ \emph {et~al.}(2018)\citenamefont
  {Lillard}, \citenamefont {Ratz}, \citenamefont {Tait},\ and\ \citenamefont
  {Trojanowski}}]{Lillard:2018zts}%
  \BibitemOpen
  \bibfield  {author} {\bibinfo {author} {\bibfnamefont {B.}~\bibnamefont
  {Lillard}}, \bibinfo {author} {\bibfnamefont {M.}~\bibnamefont {Ratz}},
  \bibinfo {author} {\bibfnamefont {M.~P.}\ \bibnamefont {Tait}, \bibfnamefont
  {Tim}}, \ and\ \bibinfo {author} {\bibfnamefont {S.}~\bibnamefont
  {Trojanowski}},\ }\href {\doibase 10.1088/1475-7516/2018/07/056} {\bibfield
  {journal} {\bibinfo  {journal} {JCAP}\ }\textbf {\bibinfo {volume} {1807}},\
  \bibinfo {pages} {056} (\bibinfo {year} {2018})},\ \Eprint
  {http://arxiv.org/abs/1804.03662} {arXiv:1804.03662 [hep-ph]} \BibitemShut
  {NoStop}%
\bibitem [{\citenamefont {B{\"o}deker}\ and\ \citenamefont
  {Schr{\"o}der}(2019)}]{Bodeker:2019ajh}%
  \BibitemOpen
  \bibfield  {author} {\bibinfo {author} {\bibfnamefont {D.}~\bibnamefont
  {B{\"o}deker}}\ and\ \bibinfo {author} {\bibfnamefont {D.}~\bibnamefont
  {Schr{\"o}der}},\ }\href@noop {} {\  (\bibinfo {year} {2019})},\ \Eprint
  {http://arxiv.org/abs/1902.07220} {arXiv:1902.07220 [hep-ph]} \BibitemShut
  {NoStop}%
\bibitem [{\citenamefont {Campbell}\ \emph {et~al.}(1992)\citenamefont
  {Campbell}, \citenamefont {Davidson}, \citenamefont {Ellis},\ and\
  \citenamefont {Olive}}]{Campbell:1992jd}%
  \BibitemOpen
  \bibfield  {author} {\bibinfo {author} {\bibfnamefont {B.~A.}\ \bibnamefont
  {Campbell}}, \bibinfo {author} {\bibfnamefont {S.}~\bibnamefont {Davidson}},
  \bibinfo {author} {\bibfnamefont {J.~R.}\ \bibnamefont {Ellis}}, \ and\
  \bibinfo {author} {\bibfnamefont {K.~A.}\ \bibnamefont {Olive}},\ }\href
  {\doibase 10.1016/0370-2693(92)91079-O} {\bibfield  {journal} {\bibinfo
  {journal} {Phys. Lett.}\ }\textbf {\bibinfo {volume} {B297}},\ \bibinfo
  {pages} {118} (\bibinfo {year} {1992})},\ \Eprint
  {http://arxiv.org/abs/hep-ph/9302221} {arXiv:hep-ph/9302221 [hep-ph]}
  \BibitemShut {NoStop}%
\bibitem [{\citenamefont {Cline}\ \emph {et~al.}(1994)\citenamefont {Cline},
  \citenamefont {Kainulainen},\ and\ \citenamefont {Olive}}]{Cline:1993bd}%
  \BibitemOpen
  \bibfield  {author} {\bibinfo {author} {\bibfnamefont {J.~M.}\ \bibnamefont
  {Cline}}, \bibinfo {author} {\bibfnamefont {K.}~\bibnamefont {Kainulainen}},
  \ and\ \bibinfo {author} {\bibfnamefont {K.~A.}\ \bibnamefont {Olive}},\
  }\href {\doibase 10.1103/PhysRevD.49.6394} {\bibfield  {journal} {\bibinfo
  {journal} {Phys. Rev.}\ }\textbf {\bibinfo {volume} {D49}},\ \bibinfo {pages}
  {6394} (\bibinfo {year} {1994})},\ \Eprint
  {http://arxiv.org/abs/hep-ph/9401208} {arXiv:hep-ph/9401208 [hep-ph]}
  \BibitemShut {NoStop}%
\bibitem [{\citenamefont {Klinkhamer}\ and\ \citenamefont
  {Manton}(1984)}]{Klinkhamer:1984di}%
  \BibitemOpen
  \bibfield  {author} {\bibinfo {author} {\bibfnamefont {F.~R.}\ \bibnamefont
  {Klinkhamer}}\ and\ \bibinfo {author} {\bibfnamefont {N.~S.}\ \bibnamefont
  {Manton}},\ }\href {\doibase 10.1103/PhysRevD.30.2212} {\bibfield  {journal}
  {\bibinfo  {journal} {Phys. Rev.}\ }\textbf {\bibinfo {volume} {D30}},\
  \bibinfo {pages} {2212} (\bibinfo {year} {1984})}\BibitemShut {NoStop}%
\bibitem [{\citenamefont {Kuzmin}\ \emph {et~al.}(1985)\citenamefont {Kuzmin},
  \citenamefont {Rubakov},\ and\ \citenamefont {Shaposhnikov}}]{Kuzmin:1985mm}%
  \BibitemOpen
  \bibfield  {author} {\bibinfo {author} {\bibfnamefont {V.~A.}\ \bibnamefont
  {Kuzmin}}, \bibinfo {author} {\bibfnamefont {V.~A.}\ \bibnamefont {Rubakov}},
  \ and\ \bibinfo {author} {\bibfnamefont {M.~E.}\ \bibnamefont
  {Shaposhnikov}},\ }\href {\doibase 10.1016/0370-2693(85)91028-7} {\bibfield
  {journal} {\bibinfo  {journal} {Phys. Lett.}\ }\textbf {\bibinfo {volume}
  {155B}},\ \bibinfo {pages} {36} (\bibinfo {year} {1985})}\BibitemShut
  {NoStop}%
\bibitem [{\citenamefont {Bauer}\ \emph {et~al.}(2016)\citenamefont {Bauer},
  \citenamefont {Schell},\ and\ \citenamefont {Plehn}}]{Bauer:2016rxs}%
  \BibitemOpen
  \bibfield  {author} {\bibinfo {author} {\bibfnamefont {M.}~\bibnamefont
  {Bauer}}, \bibinfo {author} {\bibfnamefont {T.}~\bibnamefont {Schell}}, \
  and\ \bibinfo {author} {\bibfnamefont {T.}~\bibnamefont {Plehn}},\ }\href
  {\doibase 10.1103/PhysRevD.94.056003} {\bibfield  {journal} {\bibinfo
  {journal} {Phys. Rev.}\ }\textbf {\bibinfo {volume} {D94}},\ \bibinfo {pages}
  {056003} (\bibinfo {year} {2016})},\ \Eprint
  {http://arxiv.org/abs/1603.06950} {arXiv:1603.06950 [hep-ph]} \BibitemShut
  {NoStop}%
\end{thebibliography}%

\end{document}